\documentclass{article}
\usepackage{spconf,amsmath,graphicx}
\usepackage{epsfig}
\usepackage{epstopdf}
\usepackage{subcaption}
\usepackage{titlesec}
\usepackage{cite}
\usepackage{amssymb}
\usepackage{algorithm}
\usepackage{algpseudocode}
\title{Persistent Homology of Attractors For Action Recognition}

\name{Vinay Venkataraman$^{1,2}$, Karthikeyan Natesan Ramamurthy$^{3}$, and Pavan Turaga$^{1,2}$}
\address{$^{1}$School of Arts, Media and Engineering \\ $^{2}$School of Electrical, Computer and Energy Engineering \\
	Arizona State University, Tempe, AZ \\
	$^{3}$Mathematical Sciences and Analytics Department, IBM Thomas J. Watson Research Center \\
	Yorktown Heights, NY}

\begin{document}
	%\ninept
	%
	\maketitle
	\begin{abstract}
		In this paper, we propose a novel framework for dynamical analysis of human actions from 3D motion capture data using topological data analysis. We model human actions using the topological features of the attractor of the dynamical system. We reconstruct the phase-space of time series corresponding to actions using time-delay embedding, and compute the persistent homology of the phase-space reconstruction. In order to better represent the topological properties of the phase-space, we incorporate the temporal adjacency information when computing the homology groups. The persistence of these homology groups encoded using persistence diagrams are used as features for the actions. Our experiments with action recognition using these features demonstrate that the proposed approach outperforms other baseline methods.

%In addition, we propose a novel idea of encoding temporal information in traditional topological features, such as the persistence diagrams, to better represent action data. Our experiments demonstrates that the proposed method outperforms other traditional methods.
%The key idea we propose here is to use the topological features of the attractor of the dynamical system as a feature representation, for modeling actions. 
	\end{abstract}
\section{Introduction}
\label{sec:intro}
	
The rapid technological advancements in sensing and computing has resulted in large amounts of data warranting the development of new methods for their analysis.
%which requires powerful tools to analyze. 
In the past decade, topological data analysis (TDA) has shown to be a promising new paradigm for analyzing and deriving inferences \cite{carlsson2009topology}. In this paper, we explore the suitability of TDA for analyzing human actions by modeling each action as a dynamical system and extracting the topological features of the attractor. These features are then used in a demonstrative application of classifying actions.

%and propose a framework for human activity analysis utilizing the same for applications such as action recognition. 

%methods have been investigated in various communities, and the work by Carlsson establishes that persistent homology can be used as a powerful topological data analysis approach for effectively analyzing large datasets. 

The task of recognizing human activities has a wide range of applications such as surveillance, health monitoring and animation. Modeling the spatio-temporal evolution of human body joints is traditionally accomplished by defining a state space and learning a function that maps the current state to the next state \cite{bissacco2001recognition,ralaivola2003dynamical}. An alternate approach proposed derives a representation for the dynamical system directly from the observation data using tools from chaos theory \cite{ali2007chaotic,venkataraman2013attractor,vinay_PAMI,venkataramandynamical}, thereby learning a generalized model representation suitable for analyzing a wide range of dynamical phenomenon. In this paper, we use the framework proposed in \cite{ali2007chaotic,venkataraman2013attractor} to extract a reconstructed phase-space from the available time series data, which preserves the topological properties of the underlying dynamical system of a given action. We treat the reconstructed attractor as a point cloud and we extract topological features from the point cloud based on persistent homology \cite{edelsbrunner2002topological, carlsson2014topological}.

\section{Related Work}
Human activity analysis is a well-studied problem in the vision community with extensive literature on the subject. We suggest the readers to refer \cite{aggarwal2011human,gavrila1999visual} for a detailed review of the approaches for modeling and recognition of human activities. Since our contribution in this paper is related to topological data analysis and non-parametric approaches for dynamical system analysis for action modeling, we restrict our discussion to related methods.
\vspace{-12pt}
\paragraph{Activity Analysis using Dynamical Invariants:}
Traditional methods for action recognition by parametric modeling approaches impose a model and learn the associated parameters from the training data. Hidden Markov Models (HMMs) \cite{rabiner1989tutorial} and Linear Dynamical Systems (LDSs) \cite{casti1986linear} are the most popular parametric modeling approaches employed for action recognition \cite{yamato1992recognizing,wilson1995learning,vaswani2005shape,cuntoor2007epitomic} and gait analysis \cite{kale2004identification,liu2006improved,bissacco2001recognition}. Nonlinear parametric modeling approaches like Switching Linear Dynamical Systems (SLDSs) have been utilized to model complex activities composed of sequences of short segments modeled by LDS \cite{bregler1997learning}. While, nonlinear approaches can provide a more accurate model, it is difficult to precisely learn the model parameters. In addition, one would only approximate the true-dynamics of the system with attempts to fit a model to the experimental data. An alternative nonparametric action modeling approach based on tools from chaos theory, with no assumptions on the underlying dynamical system like the largest Lyapunov exponent, correlation dimension and correlation integral, have been extensively used to model human actions \cite{ali2007chaotic,dingwell2000nonlinear,perc2005dynamics,stergiou2011human}. 
\vspace{-12pt}
\paragraph{Topological Data Analysis:}
Topological data analysis has gained its importance in analyzing point cloud data \cite{carlsson2014topological}, and is seen as a tool to obtain the \textit{shape} of high-dimensional data as opposed to geometric approaches that try to understand the \textit{size} of the data. Such tools are also very useful in visualization applications \cite{singh2007topological,du1999centroidal}. The representations of persistent homology such as persistence diagrams and barcodes have several applications, such as speech signal analysis \cite{brown2009nonlinear}, wheeze detection \cite{emrani2014persistent}, document structure representation  \cite{zhu2013persistent}, detection of cancer \cite{nanda2014simplicial}, characterizing decision surfaces in classifiers \cite{VarshneyPersistent} to name a few. There are also a number of freely available software for computing persistent homology from point clouds  \cite{tausz2011javaplex,mischaikow2013morse}.

%Similar idea of analysis of persistent homology on time-delay embeddings of human speech signals have been used by Knudson et al. to learn various statistics \cite{}. Persistent homology has also been used as a new text representation to get document structure representation in the field of natural language processing Recent work by Nanda \textit{et al.} employ persistent homology tools to perform various inference tasks in biological systems including detection of breast cancer \cite{}. Given such real-world applications and recent software tools available for free , we see persistent homology as a powerful approach suitable for extracting representative features of the shape of the reconstructed attractor from action data.

\begin{figure*}
	\centering
		\begin{subfigure}[p]{0.22\textwidth}
			\includegraphics[width=\textwidth]{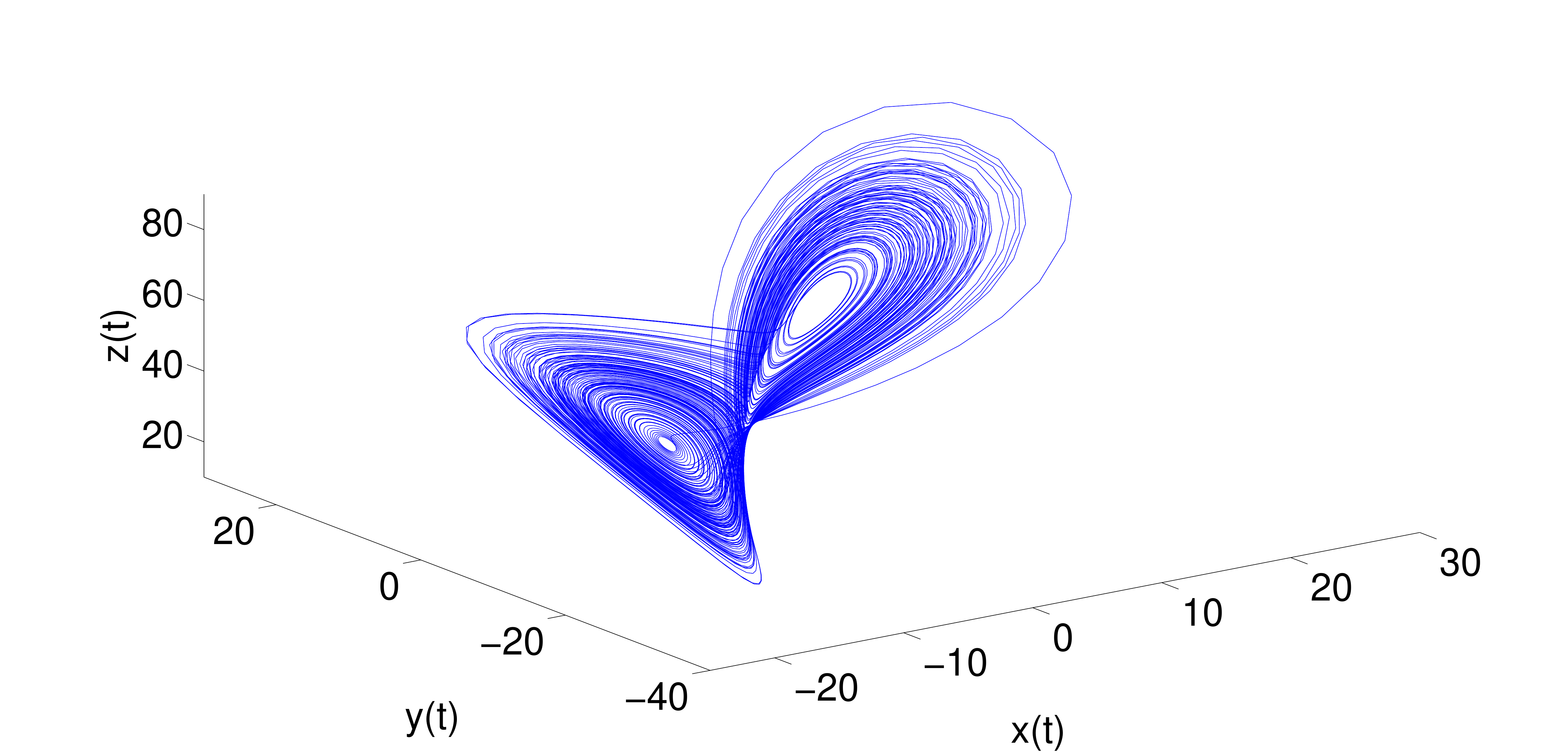}
			\caption{Lorenz Attractor}
		\end{subfigure}
	\begin{subfigure}[p]{0.22\textwidth}
		\includegraphics[width=\textwidth]{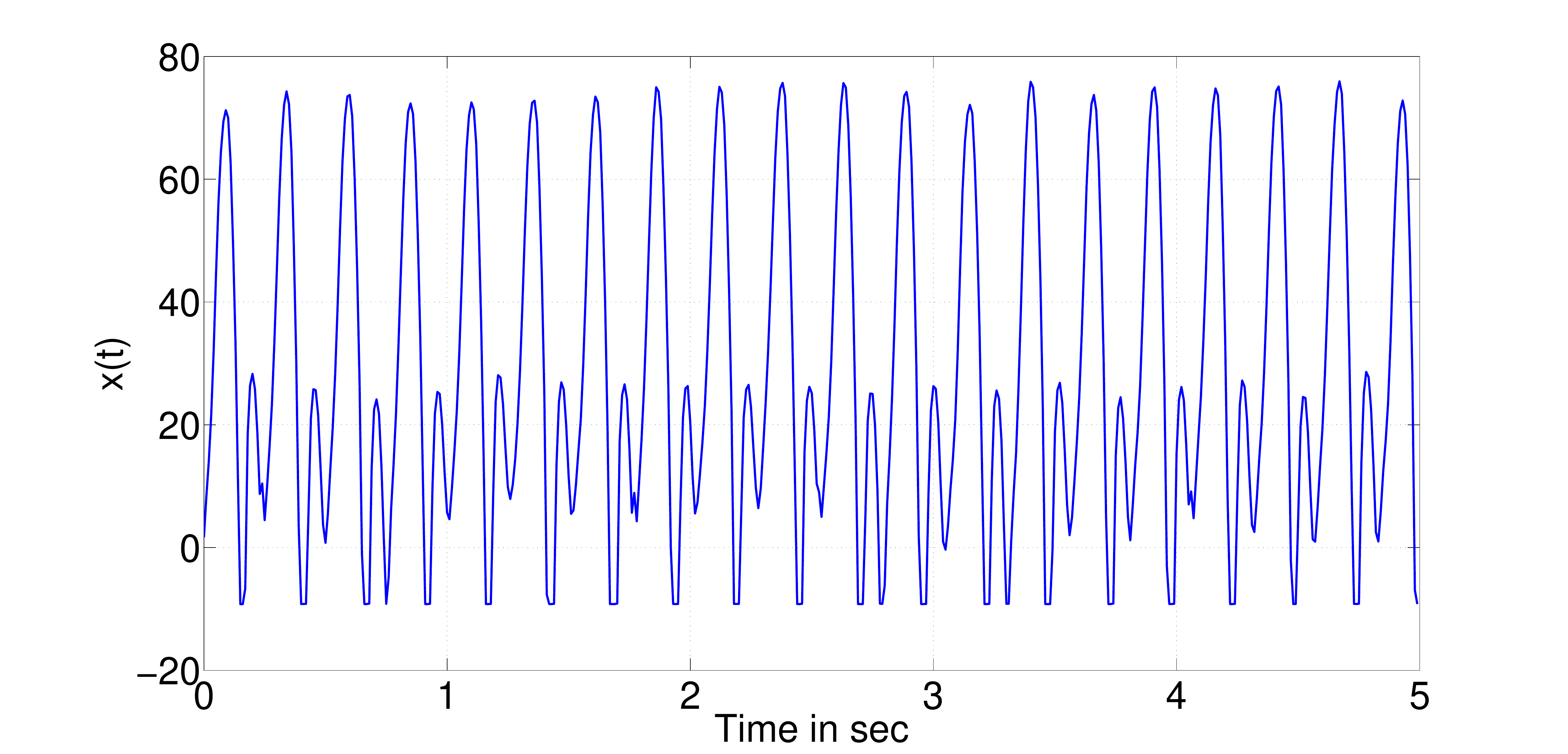}
		\caption{Time series data}
	\end{subfigure}
	\begin{subfigure}[p]{0.22\textwidth}
		\includegraphics[width=\textwidth]{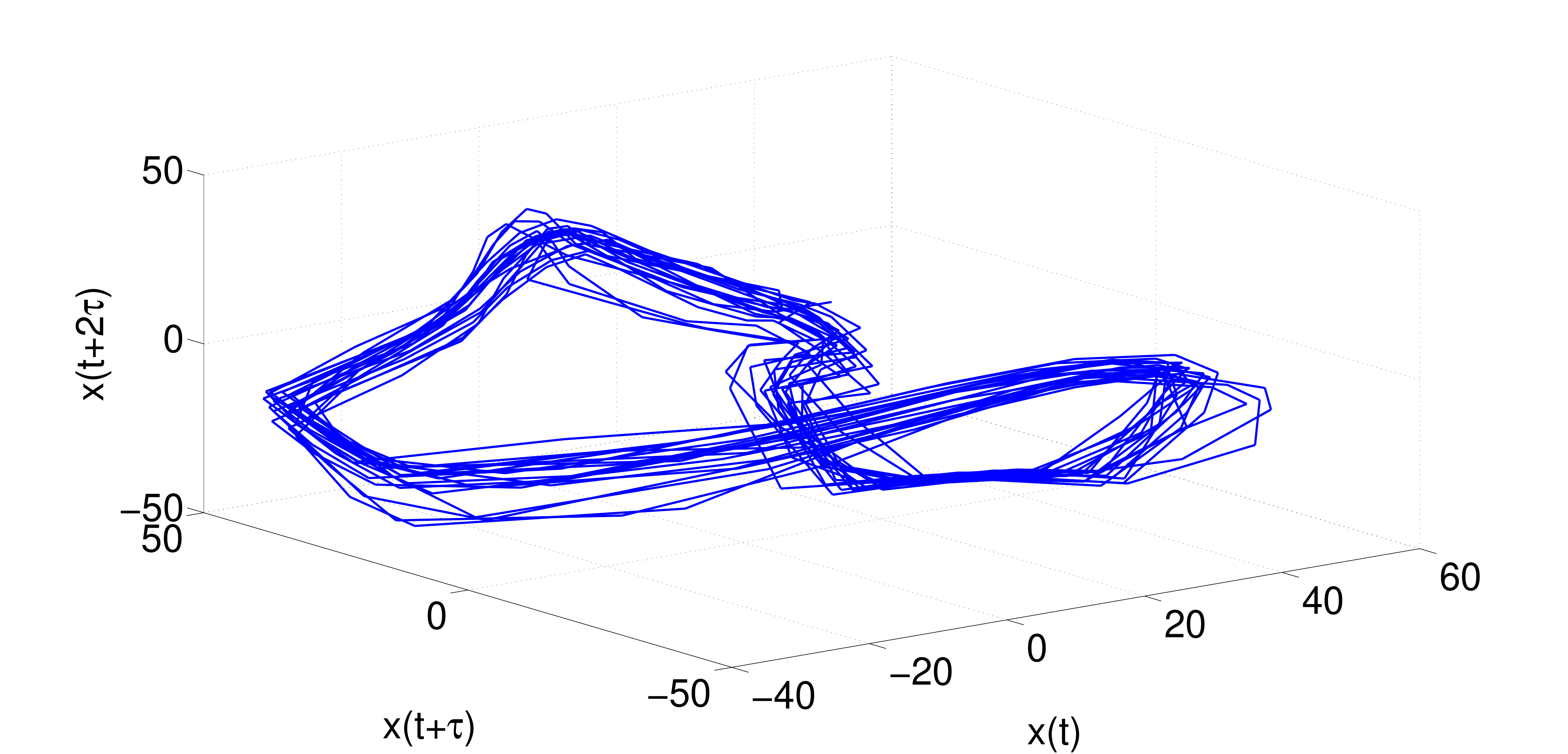}
		\caption{Reconstructed Phase Space}
	\end{subfigure}
	\begin{subfigure}[p]{0.3\textwidth}
		\includegraphics[width=\textwidth]{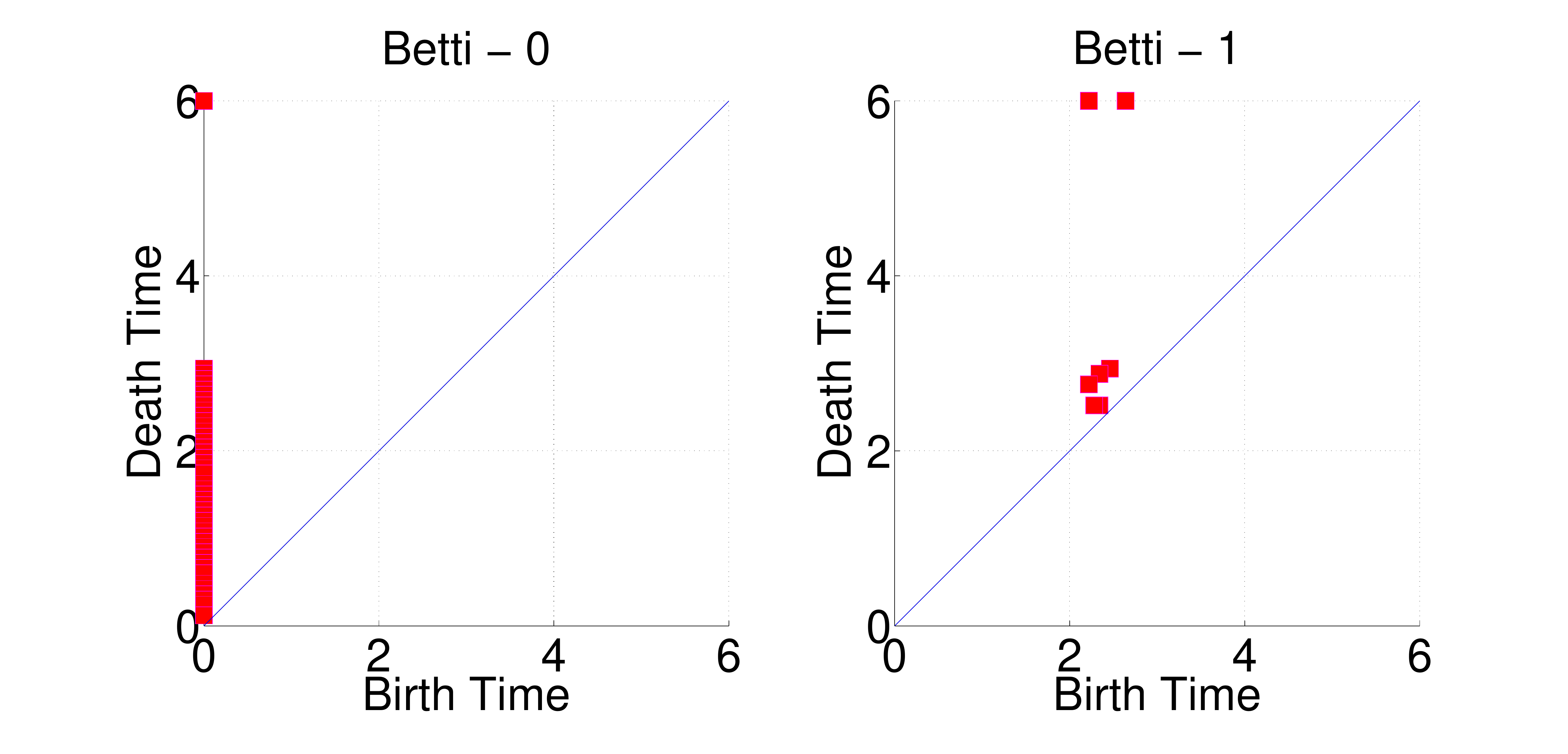}
		\caption{Persistence Diagram}
	\end{subfigure}
			\begin{subfigure}[p]{0.22\textwidth}
				\includegraphics[width=\textwidth]{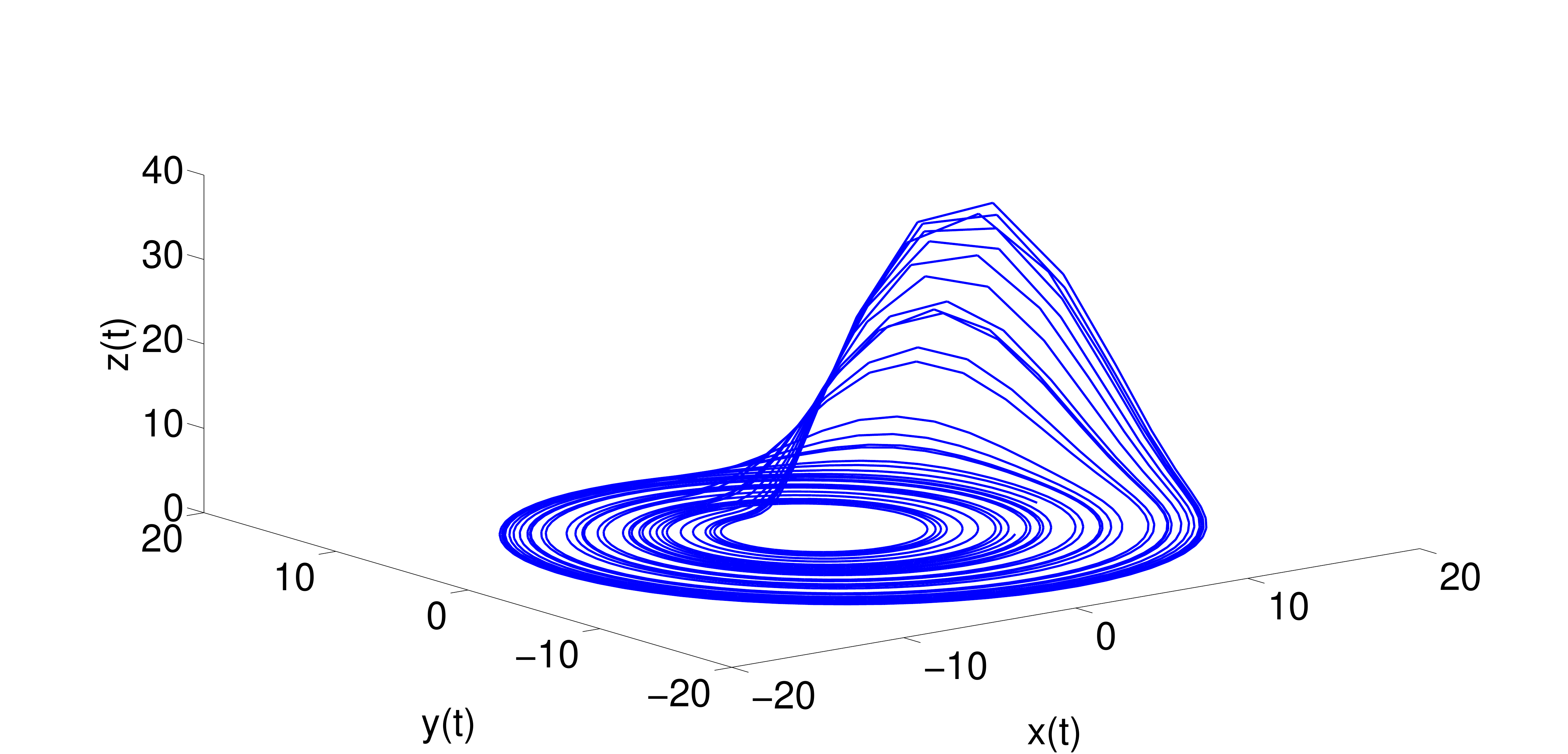}
				\caption{Rossler Attractor}
			\end{subfigure}
	\begin{subfigure}[p]{0.22\textwidth}
		\includegraphics[width=\textwidth]{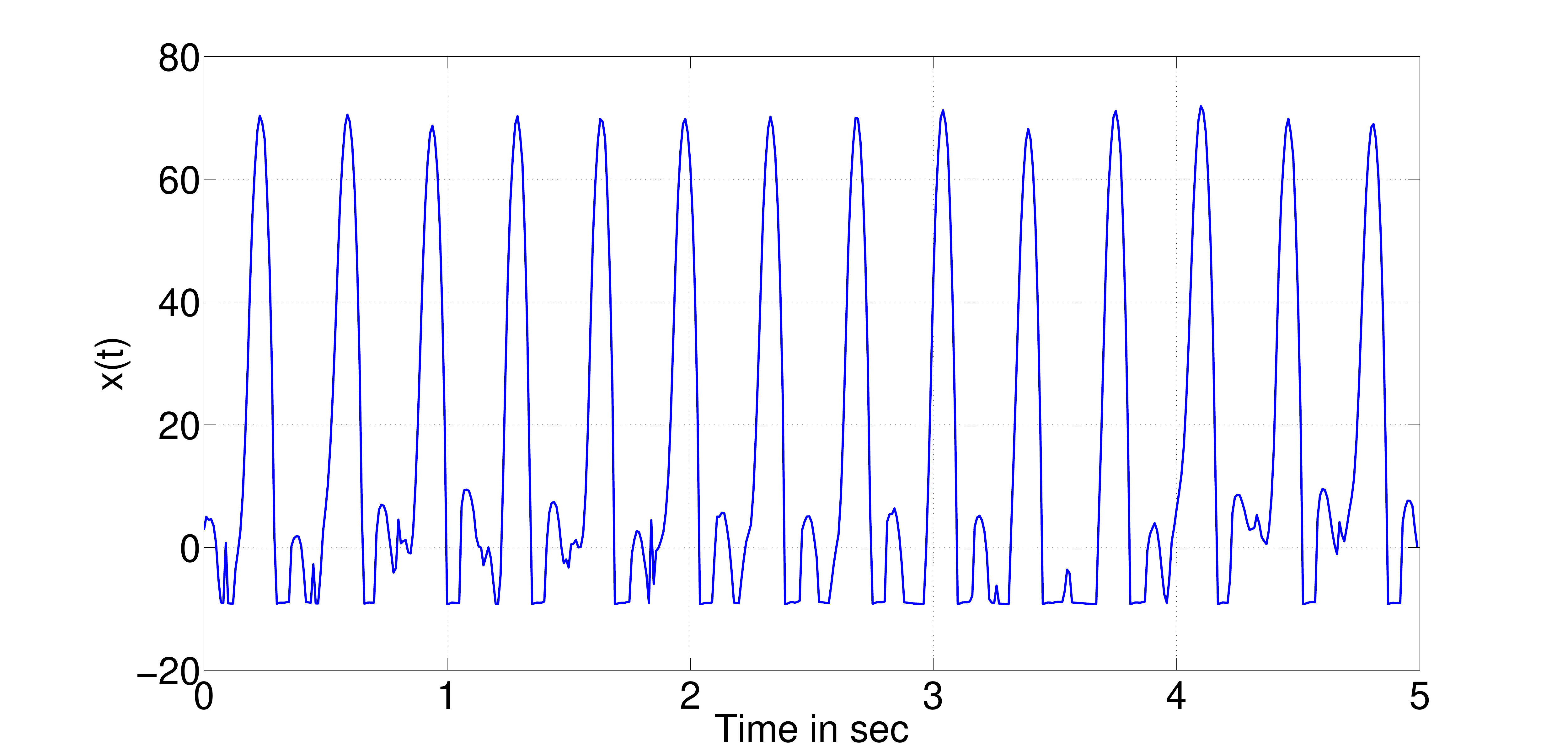}
		\caption{Time series data}
	\end{subfigure}
	\begin{subfigure}[p]{0.22\textwidth}
		\includegraphics[width=\textwidth]{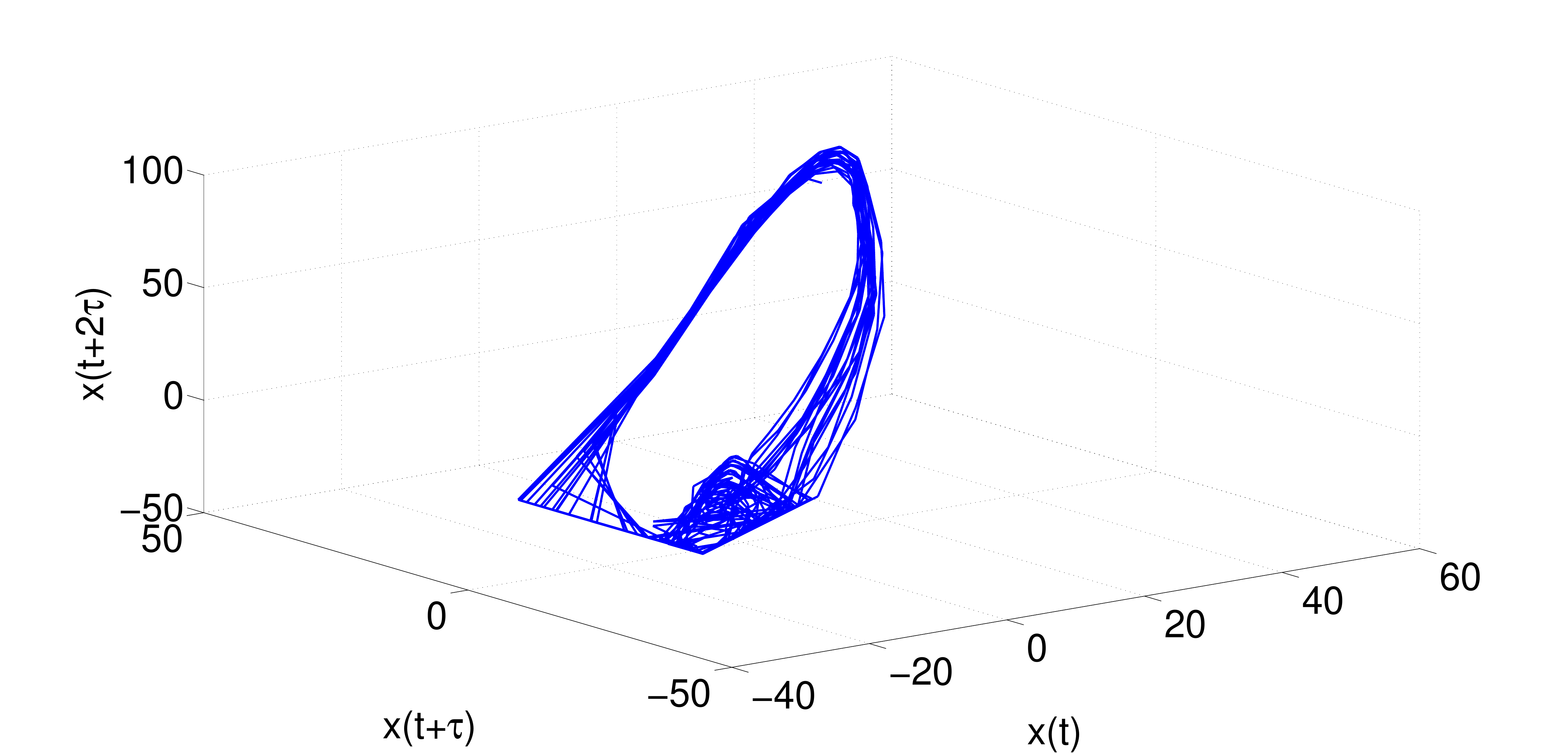}
		\caption{Reconstructed Phase Space}
	\end{subfigure}
	\begin{subfigure}[p]{0.3\textwidth}
		\includegraphics[width=\textwidth]{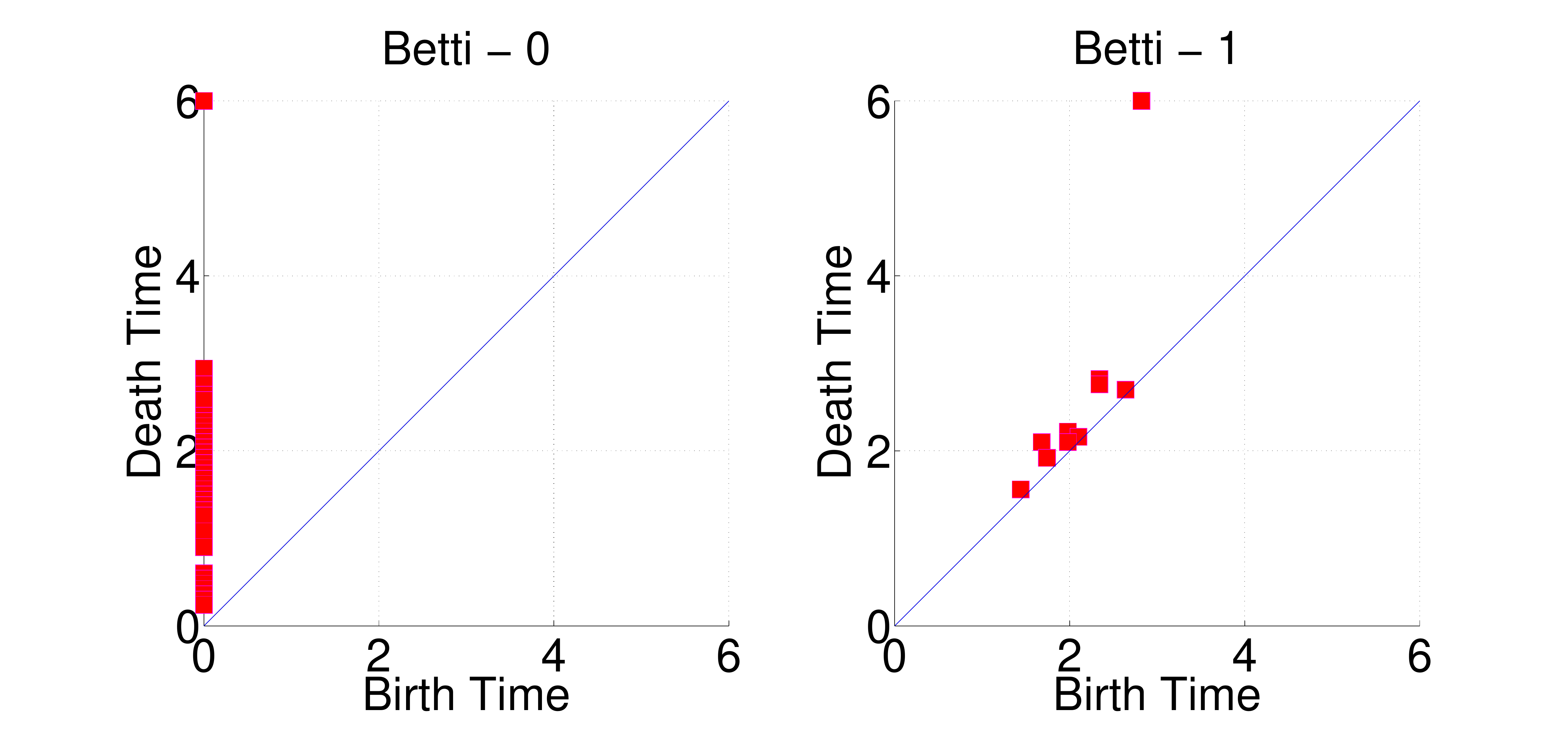}
		\caption{Persistence Diagram}
	\end{subfigure}
	\caption{Phase space reconstruction of dynamical attractors by delay embedding. (a), (e) shows the $3$D view of trajectories of Lorenz and Rossler attractors. The one-dimensional time series (observed) of the Lorenz and Rossler systems are shown in (b), (f). (c), (g) shows the reconstructed phase-space from observed time series using delay embedding. The above example illustrates that the reconstructed phase-space preserves certain topological properties of the original attractor.}
	\label{fig:PhaseSpace}
\end{figure*}	
	
\vspace{-12pt}	
\paragraph{Contributions:}	
Our work has the following contributions: (1) We treat the reconstructed phase-space of the dynamical system as a point cloud and derive features based on homological persistence. (2) We incorporate links between adjacent time points when building simplicial complexes from the point cloud. (3) We demonstrate the value of the proposed framework in an action recognition task on a publicly available motion capture dataset, using a nearest neighbor classifier with the  the persistence-based features.

%The proposedWe evaluate the proposed framework for action recognition task 

\vspace{-12pt}	
\paragraph{Outline:}
In section \ref{sec:prelimiaries}, we introduce the theoretical concepts of phase-space reconstruction and persistent homology. The feature which encodes the temporal evolution information in the persistence diagrams will be introduced in section \ref{sec:Persistencediag}. In section \ref{Experiments}, we present our experimental results on the motion capture dataset \cite{ali2007chaotic}.
	
\section{Preliminaries}
\label{sec:prelimiaries}
In this section, we introduce the background necessary to develop an understanding of nonlinear dynamical system analysis using tools from chaos theory and persistent homology.
\subsection{Phase Space Reconstruction}
The data that we obtain from sensors is usually a projection of the original dynamical system to a lower dimensional space, and hence do not represent all the variables in the system. Hence, the available data is  insufficient to model the dynamics of the system. To address this, we have to employ methods for reconstructing the attractor to obtain a phase-space which preserves the important topological properties of the original dynamical system. This process is required to find the mapping function between the one-dimensional observed time series data and the $m$-dimensional attractor, with the assumption that all variables of the system influence one another. The concept of phase-space reconstruction was proposed in the embedding theorem proposed by Takens, called Takens' embedding theorem \cite{Takens}. For a discrete dynamical system with a multidimensional phase-space, time-delay vectors (or embedding vectors) are obtained by concatenation of time-delayed samples given by 
\begin{equation}
\textbf{x}_{i}(n) = [x_{i}(n),x_{i}(n+\tau),\cdots,x_{i}(n+(m-1)\tau)]^T. 
\label{EmVec}
\end{equation}
where $m$ is the embedding dimension and $\tau$ is the embedding delay. The idea here is that for a sufficiently large $m$, the important topological properties of the unknown multidimensional system are reproduced in the reconstructed phase-space \cite{abarbanel1996analysis}. The process of phase-space reconstruction from a one-dimensional observed time-series of a Lorenz and Rossler system is shown in Fig \ref{fig:PhaseSpace}, where the reconstructed phase-space and the original attractor are topologically equivalent.

%Today's sensing systems do not allow us to observe the variables of the system, giving access to data which is usually a projection of the original dynamical system to a lower dimensional space. Hence, the available data does not represent the actual phase-space and is insufficient to model the dynamics of the system.

\subsection{Persistent Homology}
\label{sec:persistent homology}
Consider a point cloud of \textit{T} data samples in $\mathbb{R}^D$: \textbf{X} = $[\textbf{x}_{1}, \textbf{x}_{2}$, \ldots, $\textbf{x}_{T}]^T$. The point cloud data can be viewed as samples from a unknown shape. Our aim is to estimate the topological properties of the underlying shape by constructing a simplicial complex $\mathcal{S}$ using the point cloud  $\mathbf{X}$ and examining the topology of the complex. A simplicial complex is a set of simplices constructed from $\mathbf{X}$ glued together in a particular way. It is denoted by $\mathcal{S} = (\mathbf{X} , \Sigma)$, where $\Sigma$ is a family of non-empty subsets of $\mathbf{X}$, with each element  $\sigma \in \Sigma$ being a simplex. The other necessary condition is that $\sigma \in \Sigma$ and $k \subseteq \sigma$ implies that $k \in \Sigma$. The simplices are usually constructed using some neighborhood rule, such as the $\epsilon-$neighborhood, where $\epsilon$ is the scale parameter.

%A zero-dimensional simplex is a point, a one-dimensional simplex is a line segment, a two-dimensional simplex is a triangle, a three-dimensional simplex is a tetrahedron, and so on.  Specifically, a simplicial complex \textit{\textbf{S}} = (\textbf{X} , $\Sigma$), where $\Sigma$ is a family of non-empty subsets of \textbf{X} such that each subset $\sigma$ $\in$ $\Sigma$ is a simplex. Furthermore, the following condition must also hold: $\sigma$ $\in$ $\Sigma$ and $\tau \subseteq \sigma$ implies that $\tau \in \Sigma$. In forming these non-empty subsets of points that form a simplex, we only consider subsets of points that are close to each other. 

We are interested in computing the rank of homology groups of a given dimension, aka,  Betti numbers ($\beta$), since they are one of the simple but informative characterizations of topology of the point cloud. Betti$-0$ or $\beta_0$ denotes the number of connected components, $\beta_1$, the number of holes of dimension$-1$,  $\beta_2$, the number of holes of dimension$-2$ and so on. Betti numbers depend on the scale (which is same as the scale used with $\epsilon-$nearest neighbors) at which the complex is constructed. Homology groups that are stable across a wide range of scale values, i.e., \textit{persistent homology groups}, are the ones that provide the most information about the underlying shape. Homology  that do not persist are considered to be noise. The Betti numbers of a given dimension can be compactly encoded in a $2-$dimensional plot, which provides the birth versus death times of each homology group, also known as the persistence diagram. Persistence diagrams are multi-sets of points, with infinite number of points on the diagonal where birth time equals death time. They admit several metrics and hence distances between two diagrams can be estimated numerically \cite{kerbergeometry}. 

Various approaches exist for constructing simplicial complexes from $\mathbf{X}$ at a given scale $\epsilon$. In our work, we use the Vietoris-Rips (VR) complex, VR($\mathbf{X}, \epsilon$), where a simplex is created if and only if the Euclidean distance between every pair of points is less than $\epsilon$ \cite{zomorodian2010fast}. Efficient construction of the VR complex can proceed by creating an $\epsilon$-neighborhood graph, also referred to as the one-skeleton of $\mathcal{S}$. Then inductively, triplets of edges that form a triangle are taken as two-dimensional simplices, sets of four two-dimensional simplices that form a tetrahedron are taken as three-dimensional simplices, and so on. This is repeated for increasing values of scale, known as filtration, and the persistence diagrams are estimated. Although several types of topological features can be extracted from point clouds, in our work, we will use it to refer exclusively to persistence diagrams.

\begin{algorithm}[t!]
	\caption{Persistence diagrams from phase-space}
	\label{algorithm}
	\begin{algorithmic}[1]		
		\State \textbf{Input}: $\textbf{x}_{i}(n) \in \mathbb{R}^D, n = 1, \dots T $
		\State \textbf{Output}: Persistence diagram for homology group dimensions 0 \& 1.
		\For{$i = 1 \to D$ }
		\State Reconstruct attractor using method of delays \cite{abarbanel1996analysis} \par
		$\textbf{x}_{i}(n) = [x_{i}(n), x_{i}(n+\tau),\cdots,x_{i}(n+(m-1)\tau)]^T.$
		\State Construct metric space encoding temporal evolution \par
		Temporal link between $[\textbf{x}_{i}(n-1), \textbf{x}_{i}(n), \textbf{x}_{i}(n+1)].$ 
		\State Build Vietoris-Rips complexes \cite{tausz2011javaplex,zomorodian2010fast}
		\EndFor
	\end{algorithmic}
\end{algorithm}

\section{Topological Features from Attractor}
\label{sec:Persistencediag}
Although VR complexes can successfully retrieve the topological features of a general point cloud, topological features that incorporate the dynamical evolution in phase-space can model actions better. In this section, we present a method to encode temporal information in persistence diagrams which in turn can be used as a representative topological feature for the reconstructed phase-space. 

  % the topological features of the reconstructed phase-space 
 %for classification between lorenz and rossler attractors, with the number of one-dimensional holes being two and one respectively. Hence, our aim will be to extract feature representations for the topology of the reconstructed phase-space. 
%The topological tool of persistent homology discussed in section \ref{sec:persistent homology} defines the simplicial complexes at any given scale parameter $\epsilon$. 

Methods to build simplicial complexes from the point cloud data, such as the VR filtration approach, only takes into consideration the adjacency in space, but not in time. An activity is a resultant of coordinated movement of body joints and their respective interdependencies to achieve a goal-directed task with temporal information in trajectories of body joints. Modeling the underlying dynamics in the trajectories forms the core idea in designing action recognition systems. Therefore, we explicitly we create temporal links between $\mathbf{x}_{i}(n-1)$, $\mathbf{x}_{i}(n)$, and $\mathbf{x}_{i}(n+1)$ in the one-skeleton of $\mathcal{S}$, thereby creating a metric space which encodes adjacency in both space and time. The persistence diagrams for homology groups of dimensions $0$ and $1$ are then estimated. The pseudo code for our framework is outlined in algorithm \ref{algorithm}. 

As a demonstrative example, we use this approach to estimate the persistence diagrams of Lorenz and Rossler attractors. From Fig. \ref{fig:PhaseSpace}, we see that for the Lorenz attractor, the ranks of homology groups that persist are,  $\beta_0 = 1$ and $ \beta_1= 1$, whereas for the Rossler attractor,  $\beta_0 = 1$ and, $\beta_1= 2$. Clearly they indicate the connected components and $1-$dimensional holes in each of the cases. Note that the points close to the diagonal are considered to be noise with their birth and death times being close to each other. Therefore these points represent homology groups that die in a short time after they are born.

%Fig. \ref{fig:PhaseSpace} shows persistence diagrams for Lorenz and Rossler models with two and one one-dimensional holes respectively. This is clearly captured by the persistence diagrams for homology group dimension $1$ in Fig. \ref{fig:PhaseSpace}(d) \& (h), with two homology groups being \textit{persistent} for Lorenz and similarly one simplex for Rossler being \textit{persistent}. 

\vspace{-12pt}
\paragraph{Distance Between Persistence Diagrams:}
For any two persistence diagrams $X$ and $Y$, the distance between the diagrams are usually quantified using the bottleneck distance or the $q-$Wasserstein distance \cite{kerbergeometry}. In our experiments, we use the $1$-Wasserstein distance given by,
\begin{equation}
W_1(X,Y) = \inf_{\eta:X \to Y}  \sum_{x \in X} ||x-\eta(x)||_1 
\end{equation} Since each diagram contains an infinite number of points in the diagonal, this distance is computed by pairing each point in one diagram uniquely to another non-diagonal or diagonal point in the other diagram, and then computing the distance. This can be efficiently obtained with the Hungarian algorithm or using a more efficient variant \cite{kerbergeometry}.

%Persistence diagrams summarize the concept of persistent homology as two-dimensional plot with every point $(x, y)$ representing features (simplices) that appear at scale $x$ (birth time) and disappear at scale $y$ (death time). 

\section{Experimental Results}
\label{Experiments}
The proposed framework for topological data analysis for action representation was evaluated on the motion capture dataset \cite{ali2007chaotic}.

\textbf{Baseline:} To evaluate the effectiveness of our framework, we provide comparative results using $10-$dimensional feature vectors\footnote{Code available at \\ http://www.physik3.gwdg.de/tstool/HTML/index.html} of traditional chaotic invariants obtained by concatenating the largest Lyapunov exponent, correlation dimension and correlation integral (for $8$ values of radius). The results with this approach are denoted with \textit{Chaos} in Table \ref{tab:results1}. We also tabulate the results using persistence diagrams obtained from VR filtrations with no additional temporal encoding (\textit{VR Complex}), and a recent shape-theoretic framework \textbf{D2} and \textbf{DT2} \cite{vinay_PAMI}. The evaluation with VR complexes follow the same protocol as our proposed approach described below.

%as a baseline to validate our assumption that encoding temporal information in the persistance diagrams will provide improvement in action recognition. We also compare our performance with . 

\subsection{Motion Capture Data}
We evaluate the performance of the proposed framework using $3$-dimensional motion capture sequences of body joints used in the \cite{ali2007chaotic}. The dataset is a collection of five actions: \textit{dance, jump, run, sit} and \textit{walk} with $31, 14, 30, 35$ and $48$ instances respectively. The dataset provides $3-$dimensional time-series from $17$ body joints which were further divided into scalar time-series resulting in a $51$-dimensional vector representation for each action.  We generate $100$ random splits having $5$ testing examples from each action class and use a nearest neighbor classifier with the $1-$Wasserstein distance measure. The mean recognition rates for the different methods are given in Table \ref{tab:results1}. Traditional chaotic invariants (\textit{Chaos}) only achieves a mean recognition rate of $52.44\%$. The best classification performance reported on the dataset uses \textbf{DT2} dynamical shape feature achieves a mean recognition rate of $93.92\%$ which encodes temporal information. In comparison, our proposed method achieves $96.48\%$ which is significantly better than the results achieved by any of the previous methods. Clearly, topological persistence features are informative, since they summarize the feature evolution over a range of scale values when compared to chaotic invariants such as largest Lyapunov exponents. The standard deviation of classification accuracy over the different random splits are also tabulated. The class confusion matrix for the proposed framework is shown in Table \ref{tab:confTable}. 

\begin{table}[t]
	%\footnotesize
	\begin{center}
		\caption{Comparison of classification rates for different methods using nearest neighbor classifier on the motion capture dataset.}
		\begin{tabular}{|c|c|c|}
			\hline
			\textbf{Method} & \textbf{Mean Accuracy (\%)} & \textbf{Std. dev} 	\\ \hline \hline			
			Chaos \cite{ali2007chaotic} & 52.44 &	0.0081 \\ 
			VR Complex \cite{tausz2011javaplex} & 93.68 & 0.0054 \\ 
			D2 \cite{venkataraman2013attractor} & 91.96 &	0.0036	\\ 					
			DT2 \cite{vinay_PAMI} & 93.92 &	0.0051	\\ \hline \hline
			\textbf{Proposed} & \textbf{96.48} & \textbf{0.0053}\\ \hline
		\end{tabular}
		\label{tab:results1}
	\end{center}
	%\vspace{-15pt}
\end{table}

\begin{table}
	\begin{center}
		\caption{Confusion table for motion capture dataset using our proposed framework which achieves mean classification rate of $96.48\%$.}
		\begin{tabular}{| c | c | c | c | c | c |}
			\hline
			\textbf{\textit{Action}} & \textbf{Dance} & \textbf{Jump} & \textbf{Run} & \textbf{Sit} & \textbf{Walk}   	\\ \hline	\hline
			\textbf{Dance} & \textbf{0.98} & 0 & 0 & 0.02 & 0 	\\ \hline
			\textbf{Jump} & 0.08 & \textbf{0.92} & 0 & 0 & 0 		\\ \hline
			\textbf{Run} & 0 & 0 & \textbf{0.96} & 0 & 0.04 		\\ \hline
			\textbf{Sit} & 0.03 & 0 & 0 & \textbf{0.97} & 0		\\ \hline
			\textbf{Walk} & 0 & 0 & 0.01 & 0 & \textbf{0.99}		\\ \hline
		\end{tabular}  
		\label{tab:confTable}
	\end{center}
	%\vspace{-15pt}
\end{table}

\section{Conclusion and Future Work}
In this paper, we have proposed a novel topological feature representation for persistent homology which encodes temporal information in any given point cloud suitable for applications in action recognition. The proposed framework addresses the drawbacks of conventional methods, by combining the principles from nonlinear time-series analysis and topological data analysis, to extract robust and discriminative features from the reconstructed phase-space. 

%Topological data analysis of time-delay embeddings can be seen as a natural way to understand and visualize time-series data

%particularly useful to understand the periodicities of a 1D signal in a high-dimensional space in case of human activities as validated by the experimental evaluation on the motion capture dataset.
Since computing distances between persistence diagrams is similar to obtaining Wasserstein distance between two probability mass functions, a well-designed multi-resolution approach can be used to reduce complexity, particularly in applications where we only need approximate distances. Further, using recently proposed persistence kernels \cite{reininghaus2014stable} can significantly widen the scope of applications of topological persistence features.

%Computing distances between persistence diagrams using existing distance metrics is computationally very expensive. We focus our future work towards using simplifications such as the witness complexes \cite{de2004topological}. 
	
\bibliographystyle{IEEEbib}
\footnotesize
\bibliography{egbib,strings}
	
\end{document}